\documentclass[fleqn,usenatbib]{hofmann}

\usepackage{newtxtext,newtxmath}

\usepackage{float}
\restylefloat{figure}
\restylefloat{table}
\usepackage{tabularx}

\usepackage[paper=a4paper,left=17mm,right=17mm,top=20mm,bottom=20mm]{geometry}

\usepackage{caption}
\usepackage{setspace}

\usepackage{natbib}

\usepackage{color,soul}

\usepackage{hyperref}
\hypersetup{
    colorlinks=true,
    linkcolor=ultramarine3,
    filecolor=ultramarine3,      
    urlcolor=ultramarine3,
    pdftitle=Comparison of LCDM and SU(2)CMB,
    pdfpagemode=FullScreen,
    citecolor=ultramarine3
    }

\usepackage{xcolor}
\definecolor{ultramarine}{RGB}{0,32,96}
\definecolor{ultramarine2}{RGB}{0,68,204}
\definecolor{ultramarine3}{RGB}{0,0,180}

\usepackage[T1]{fontenc}

\DeclareRobustCommand{\VAN}[3]{#2}
\let\VANthebibliography\thebibliography
\def\thebibliography{\DeclareRobustCommand{\VAN}[3]{##3}\VANthebibliography}

\usepackage{graphicx,subfigure}	
\usepackage{amsmath}	
\usepackage{amsmath,array,graphicx}
\usepackage{kantlipsum}
\usepackage{hyperref}

\newcommand{\eqb}{\begin{equation}}
\newcommand{\eqe}{\end{equation}}
\newcommand{\dmb}{\begin{displaymath}}
\newcommand{\dme}{\end{displaymath}}

\newcommand{\eab}{\begin{eqnarray}}
\newcommand{\eae}{\end{eqnarray}}

\newcommand{\be}{\begin{equation}}
\newcommand{\ee}{\end{equation}}

\newcommand{\nn}{\newline}
\newcommand{\LC}{\textnormal{\textsc{l}}}

\newcommand{\SU}{\textnormal{\textsc{ym}}}
\newcommand{\YM}{\textnormal{\textsc{ym}}}

\title[Comparison of $\Lambda$CDM and SU(2)$_{\rm CMB}$]{Cosmological parameters from Planck data in SU(2)$_{\rm CMB}$, their local $\Lambda$CDM values, and the modified photon Boltzmann equation}

\author[R. Hofmann et al.]{
Ralf Hofmann$^{1,}$\thanks{E-mail: R.Hofmann@ThPhys.Uni-Heidelberg.de}\href{https://orcid.org/0000-0001-6365-0631}{\hspace{0.1mm}\includegraphics[scale=0.06]{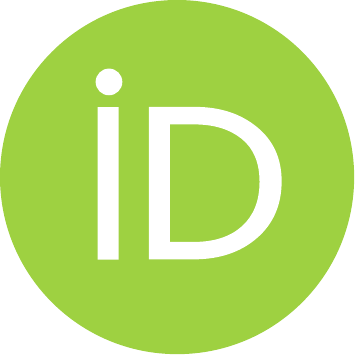}},
Janning Meinert$^{1,2}$\href{https://orcid.org/0000-0001-7582-3456}{\hspace{0.1mm}\includegraphics[scale=0.06]{orcid.pdf}}, 
and Shyam Sunder Balaji$^{1}$\vspace{1mm}
\\
$^{1}$Institut f\"ur Theoretische Physik, Universit\"at Heidelberg Philosophenweg 12, D-69120 Heidelberg, Germany\\
$^{2}$Department of Physics, Bergische Universit\"at Wuppertal, Gaußstraße 20, D-42119 Wuppertal, Germany
}

\begin{document}
\label{firstpage}
\pagerange{\pageref{firstpage}--\pageref{lastpage}}
\maketitle


\begin{abstract}
A review of the spatially flat cosmological model SU(2)$_{\rm CMB}$, minimally induced by the postulate that the Cosmic Microwave Background (CMB) is subject to an SU(2) rather than a U(1) gauge principle, is given. Cosmological parameter values, 
which are determined from the Planck CMB power spectra at small angular scales, are compared to their values in spatially flat $\Lambda$CDM 
from both local and global extractions. As a global model 
SU(2)$_{\rm CMB}$ leans towards local $\Lambda$CDM cosmology and is in tension with 
some global $\Lambda$CDM parameter values. We present spectral 
antiscreening\,/\,screening effects in SU(2)$_{\rm CMB}$ radiance within the Rayleigh-Jeans regime 
in dependence on temperature and frequency. Such radiance 
anomalies can cause CMB large-angle anomalies. Therefore, it is 
pointed out how SU(2)$_{\rm CMB}$ modifies the Boltzmann equation 
for the perturbations of the photon phase space distribution at 
low redshift and why this requires to the solve the $\ell$-hierarchy 
on a comoving momentum grid ($q$-grid) for all $z$.
\end{abstract}




\section{Introduction}

Our present age witnesses a promising change in paradigm on how to 
model and analyse the composition and dynamics of the Cosmos. 
This shift is concerned with a departure from perturbative towards nonperturbative approaches. 

Within flat $\Lambda$CDM one example on the modelling side is that nonlinear clustering observables (e.g., the galaxy-halo connection model) on cosmologically small comoving length scales (a few to tens of $h^{-1}$Mpc), which evolve out of adiabatic, Gaussian initial perturbations, 
not only are addressed by mild multiplicative deformations of 
their perturbatively evolved versions \cite{Sugiyama:2021axw} 
but by nonperturbative, high-resolution N-body simulations \cite{Miyatake:2021sdd,Miyatake:2020uhg}. 
In contrast to the former the latter method does not require an anchoring in 
high-z observables, which rely on a specific cosmological model, 
is valid if scales are not too small \cite{Miyatake:2020uhg}, and exhibits 
large signal-to-noise ratios in weak lensing signals \cite{Miyatake:2021sdd}. 

An example on the theoretical side is deconfining SU(2) Yang-Mills thermodynamics with an a priori estimate of the 
thermal ground state based on selfdual, topologically nontrivial gauge-field 
configurations \cite{bookHofmann}. Relying on this result, a postulate on the CMB being subject 
to an SU(2) rather than a U(1) gauge principle can be made, henceforth referred to as SU(2)$_{\rm CMB}$, with its Yang-Mills scale (or critical temperature $T_c$ 
for the deconfining-preconfining transition) fixed by CMB radio-frequency observations 
\cite{Fixsen:2009xn,Hofmann:2009yh}. 

The flat $\Lambda$CDM model is a minimal and successful framework to 
accommodate a wealth of cosmological data \cite{Riess_1998,Perlmutter:1998np,BOSS:2016wmc,WMAP:2003ivt}. 
Throughout the last decade, however, tensions were uncovered in certain parameter values of this model when determined by data referring to local vs. global cosmology, see \cite{Abdalla:2022yfr} for a recent, comprehensive review. Most profoundly, there is a Hubble crisis. This is  expressed by a $\sim $5$\,\sigma$ discrepancy 
between the value $H_0\sim 73.5\,$km\,s$^{-1}$\,Mpc$^{-1}$ (errors ranging between 1 and 2.5 km\,s$^{-1}$\,Mpc$^{-1}$) 
as extracted from the Hubble diagram in local, flat $\Lambda$CDM, see e.g. \cite{Riess:2021jrx}, using calibrated 
standard candles, or from strong-lensing time delays (cosmography, only astrophysics model dependence), see \cite{Wong:2019kwg}, and $H_0\sim (67.27\pm 0.60)\,$km\,s$^{-1}$\,Mpc$^{-1}$ fitted 
to CMB two-point power spectra by the Planck collaboration \cite{Planck:2018vyg} with similarly low values obtained from BAO (standard ruler) data \cite{BOSS:2016wmc} assuming flat $\Lambda$CDM to be valid globally. Next, global fits of flat $\Lambda$CDM and BBN 
yield a baryon density which is by a factor $\sim$\,3/2 higher than the value observed by direct baryon census, see e.g. \cite{Planck:2018vyg,Kirkman:2003uv} for the former and \cite{Shull:2011aa,2019} for the latter claim.  Moreover, within flat $\Lambda$CDM weak gravitational lensing effects persistently indicate a value of the clustering amplitude \cite{Miyatake:2021sdd,DES:2021wwk,Heymans:2020gsg} which relates to a value of $\sigma_8$ being by 2-3\,$\sigma$ lower compared to the value 
extracted from CMB observation \cite{Planck:2018vyg, Nunes_2021}.
Also, there is a mild tendency for an increase of 
$\Omega_m$ (by a maximum significance of $\sim 1\,\sigma$ in \cite{Miyatake:2021sdd}) compared to 
the CMB extraction in \cite{Planck:2018vyg}. Finally, we point out a $\sim$2\,$\sigma$ tension in the 
redshift $z_{\rm re}$ for reionisation between direct observation using the Gunn-Peterson trough 
\cite{Becker_2001} and the latest extraction from the Planck data \cite{Planck:2018vyg}. 
In addition to these anomalies in flat $\Lambda$CDM parameter values, there are large-angle anomalies in the CMB, 
hinting at a dynamical breaking of statistical isotropy relevant to these angular scales \cite{Tegmark:2003ve,Gordon:2005ai,Copi:2005ff,Hofmann:2013rna}.
These anomalies can be distinguished as follows: lack of large-angle CMB temperature correlation, hemispherical power asymmetry, octopole planarity and alignment with the quadrupole, point-parity anomaly, variation in cosmological parameters over the sky, and cold spot. For a comprehensive, very recent summary see \cite{Abdalla:2022yfr}. 

The purpose of the present paper is twofold. In the first half we explain the cosmological model implied by SU(2)$_{\rm CMB}$ and its fit to Planck data in \cite{Hahn:2018dih}, in particular focusing on the dark-sector parametrisation and a 
physical realisation thereof proposed in \cite{MeinertHofmann2021}. Flat $\Lambda$CDM emerges at low redshifts in this model, and we compare the according parameter values with those of recent weak-lensing and galaxy clustering analyses, local Hubble-diagram fits, observations of the onset of the epoch of reionisation by the detection of the Gunn-Peterson trough in the spectra of distant quasars, and direct baryon censuses. This is confronted with the extraction of flat $\Lambda$CDM parameters in global cosmology probes (CMB and BAO). As a result, we see a tendency that $\Omega_m$ is increased and $\sigma_8$ decreased compared to these global fits. In particular, the latest results on weak-lensing galaxy-galaxy correlation using the HSC Y1 and SDSS data yield coinciding central values of these two parameters with those of the model in \cite{Hahn:2018dih}, albeit the significance of $\Omega_m$'s deviation only is 1\,$\sigma$. Moreover, the model in \cite{Hahn:2018dih} obtains values of other cosmological parameters which point towards the values extracted from local probes, most noticeably the value of $H_0\sim (74.24\pm 1.46)\,$km\,s$^{-1}$\,Mpc$^{-1}$ deviates by less than 1\,$\sigma$ from that of \cite{Riess:2021jrx}. The latter, in turn, deviates from global extractions in flat $\Lambda$CDM by more than 5\,$\sigma$. In the second half of the paper we revisit the modification of the conventional Planck spectrum of blackbody radiance at low frequencies and temperatures with the intention to eventually implement this spectral anomaly into a particular CMB Boltzmann solver -- CLASS \cite{Ma_1995,Lesgourgues:2011re,Blas:2011rf}. We suppose \cite{Szopa:2007wy,Ludescher:2009my,Hofmann:2013rna} that a proper implementation of the according comoving energy-momentum relation in such a code conveys some of the above mentioned 
large-angle anomalies \cite{Tegmark:2003ve,Copi:2005ff} even though the projection onto $C_\ell$'s assumes statistical isotropy. 
Presently, we face technical problems in the implementation, however. This concerns the introduction of a grid in comoving momentum $q$ for the photon Boltzmann hierarchy. Therefore, no results on the low-$\ell$ CMB angular power spectra are presented here\footnote{Note that the information residing in the $C_\ell$'s is just a projection of the isotropy breaking effect since their computation assumes statistical isotropy. In \cite{Tegmark:2003ve,Copi:2005ff,Vielva:2010ng} for example, statistics are considered which measure the breaking of statistical isotropy without such a projection.}. Hoping that experts in the CMB modelling community can be interested in overcoming these problems in a 
reasonable amount of time, desirably in collaboration with the present authors, we provide the required comoving photon dispersion law in SU(2)$_{\rm CMB}$. 

This paper is organised as follows. In Sec.\,2 we review the minimal, spatially flat 
cosmological model SU(2)$_{\rm CMB}$, as it was employed in \cite{Hahn:2018dih} in fits to 2015 Planck data. We also discuss dark-sector physics, based on ultralight Planck-scale axion species 
\cite{MeinertHofmann2021}, which the minimal dark sector of SU(2)$_{\rm CMB}$ in \cite{Hahn:2018dih} may be mimicking. 
Cosmological parameter values extracted in \cite{Hahn:2018dih} are compared with global and recent local extractions within 
flat $\Lambda$CDM or by cosmography to point out a tendency of SU(2)$_{\rm CMB}$ as a global model leaning towards local, flat $\Lambda$CDM. Sec.\,3 first provides a brief review of large-angle anomalies in the CMB, based on analyses of the two satellite mission WMAP and Planck. The radiatively induced antiscreening\,/\,screening effects in the Rayleigh-Jeans regime, which are described by the screening function $G$ for the thermal SU(2)$_{\rm CMB}$ photon, could explain the CMB large-angle anomalies, see \cite{Ludescher:2009my,Hofmann:2013rna}. Therefore, we review this blackbody anomaly of spectra radiance both as a function of temperature and frequency. It is pointed out that the maximal deviation between U(1) and SU(2)$_{\rm CMB}$ radiances is constantly feeble at temperatures 
considerably larger than $T_c\text{\;$=$\;}T_0\text{\;$=$\;}2.725$\,K, rendering its detection at high temperatures, 
say $T\text{\;$=$\;}300\,$K, experimentally challenging. Next, we discuss the effects of screening function $G$ on the 
Boltzmann equation for the cosmological evolution of linear perturbations of photon phase-space distribution 
in conformal Newtonian gauge. Since low-redshift (low-$z$) photons suffer antiscreening\,/\,screening a nontrivial comoving energy-momentum 
relation persists, exhibiting a dependence on conformal time $\tau$. 
Moreover, a match between high-$z$ and low-$z$ evolution needs to be made when solving the Boltzmann hierarchy on a comoving momentum grid ($q$-grid) for {\sl all} $\tau$, including active Thomson scattering.
Finally, we point out which modules of the Boltzmann code CLASS are affected by the modified cosmological model SU(2)$_{\rm CMB}$ to simultaneously address the CMB power spectra at high-$\ell$ for cosmological parameter extraction and at low-$\ell$ to mitigate the discrepancy of TT power seen in \cite{Hahn:2018dih}. Such a lowering of TT power would be a smoking gun for the breaking of statistical isotropy at low redshift mediated by SU(2)$_{\rm CMB}$.

\section{Present status of \texorpdfstring{SU(2)$_{\rm CMB}$}{SU(2)CMB}\label{pssu2}}
\subsection{\texorpdfstring{$T$-$z$}{T-z} relation and other implications for the cosmological model\label{Tzoi}}

The change due to SU(2)$_{\rm CMB}$ in spatially flat FLRW cosmology, which, as a background model, 
appreciably starts deforming $\Lambda$CDM at redshifts well within the dark ages, is induced by a modified CMB temperature ($T$) - redshift ($z$) relation. For the reader's convenience we repeat here 
the arguments put forward in \cite{Hahn:2018dih} how this modification comes about and what it implies.  

\begin{figure}
\centering
\caption{\protect{\label{Fig-1}} Plot of function ${\cal S}(z)$ of Eq.\,(\ref{devlinS}), defined as a (multiplicative) deviation from the linear $T$-$z$ relation of 
Eq.\,(\ref{TzT>>T0}). The curvature in ${\cal S}(z)$ at 
low $z$ indicates the breaking of conformal invariance in the deconfining SU(2) Yang-Mills plasma for $T\sim T_0$ with a rapid approach towards
$\left({1}/{4}\right)^{1/3}\approx 0.63$ as $z$ increases. 
Figure adapted from \protect\cite{Hahn:2018dih}.}
\includegraphics[width=\columnwidth]{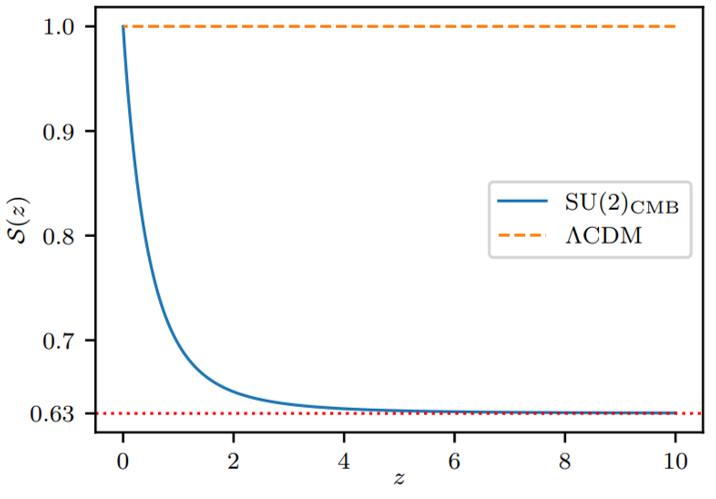} 
\end{figure}

In an (energy conserving) FLRW 
universe one demands 
\eqb
\label{enercon}
\frac{\mbox{d}\rho_\SU}{\mbox{d}a}\text{\;$=$\;}-\frac{3}{a}\left(\rho_\SU+P_\SU\right)\,,
\eqe
where $\rho_\SU$ and $P_\SU$ denote energy density and pressure, respectively, in the deconfining 
phase of SU(2) Yang-Mills thermodynamics (subscript \textsc{ym}), and $a$ refers to 
the cosmological scale factor, normalised to $a(T_0)\text{\;$=$\;}1$, where $T_c\text{\;$=$\;}T_0\text{\;$=$\;}2.725\,$K indicates the present baseline temperature of the CMB \cite{Mather:1990tfx}, interpreted as the critical temperature $T_c$ for the deconfining-preconfining phase transition in  \cite{Hofmann:2009yh}. The solution of Eq.\,(\ref{enercon}) can be recast as 
\begin{equation}\label{solt>t0}
a \equiv \frac{1}{z+1}=\exp\left(-\frac{1}{3}\log \left(\frac{s_\SU(T)}{s_\SU(T_0)}\right) \right)\,.
\end{equation}
Here the entropy density $s_\SU$ is given as 
\eqb
\label{entropydens}
s_\SU\equiv\frac{\rho_\SU+P_\SU}{T}
\eqe
which shows that the a priori estimates of the thermal ground-state contributions to pressure and 
energy density do not contribute to Eq.\,(\ref{solt>t0}). 
For large temperatures, $T\text{\;$\gg$\;}T_0$, Eq.\,(\ref{solt>t0}) can be simplified \cite{Hahn:2017yei} as 
\eqb
\label{TzT>>T0}
\frac{T(z)}{T_0}\text{\;$=$\;}\left(1/4\right)^{1/3}\,(z+1)\approx 0.63\,(z+1)\,.
\eqe 
The basis $1/4$ is the ratio between the number $n_P$ of relativistic degrees of freedom in constituting the gauge-field excitations of the plasma at $T_0$ ($n_P\text{\;$=$\;}2$) and for $T\text{\;$\gg$\;}T_0$ ($n_P\text{\;$=$\;}8$). For temperatures $T$ not much higher than $T_0$ linearity in the $T$-$z$ relation is violated 
by the Yang-Mills scale $\Lambda$ (related to $T_0$ by 
$\Lambda$\;$=$\;${2\pi T_0}/{13.87}$
\cite{bookHofmann}) breaking conformal invariance. Therefore, we define the multiplicative deviation $S(z)$ 
from linear scaling at any given temperature $T$ in the deconfining phase as
\eqb
\label{devlinS}
{\cal S}(z)\text{\;$=$\;}\left(\frac{\rho_\SU(z\text{\;$=$\;}0)+P_\SU(z\text{\;$=$\;}0)}{\rho_\SU(z)+P_\SU(z)}\frac{T^4(z)}{T^4_0}\right)^{1/3}\,.
\eqe 
As a result, the $T$-$z$ relation assumes the generally valid form
\eqb
\label{TzT}
\frac{T(z)}{T_0}\text{\;$=$\;}{\cal S}(z)\,(z+1)\,\ \ \ \ (T\text{\;$\ge$\;}T_0)\,.
\eqe
Fig.\,\ref{Fig-1} depicts function ${\cal S}(z)$.             
For the conformally invariant Yang-Mills gas and for $T\text{\;$\gg$\;}T_0$, when all eight gauge modes are nearly massless \footnote{Two polarisations for the massless mode, three polarisations for each of the two massive modes.}, the $z$ dependence of the deconfining Yang-Mills energy density $\rho_{{\SU}}$ is implied by Eq.\,(\ref{TzT>>T0}) to be 
\eqb
\label{rhosu2}
\rho_{{\SU}}(z)\text{\;$=$\;}4\,\left(\frac{1}{4}\right)^{4/3}\rho_\gamma(z)\text{\;$=$\;}
\left(\frac{1}{4}\right)^{1/3}\rho_\gamma(z)\quad(z\text{\;$\gg$\;}1)\,.
\eqe             
Here, $\rho_\gamma$ denotes the energy density of a thermal photon gas, using the U(1) $T$-$z$ relation $T\text{\;$=$\;}T_0(z+1)$. 
Again, for low temperatures conformal invariance is broken, and Eq.\,(\ref{rhosu2}) needs to be modified accordingly, see \cite{Hahn:2018dih}. For the $z$ dependence of the energy density of massless neutrinos one has for $T\text{\;$\gg$\;}T_0$
\begin{equation}\label{eq:def:omegaNu}
\Omega_\nu(z)\text{\;$=$\;}\frac{7}{8} N_{\rm eff}\,\left(\frac{16}{23} \right)^{{4}/{3}} \Omega_{\YM, \gamma}  (z) \,.
\end{equation}
In Eq.\,(\ref{eq:def:omegaNu}) a modified factor for the conversion of neutrino to CMB temperature occurs 
because of additional relativistic degrees of freedom 
during $e^+e^-$ annihilation \cite{Hofmann:2014lka}, $\Omega_{\YM, \gamma}(z)$ refers to the photon part of the 
density parameter in deconfining SU(2)$_{\text{CMB}}$ thermodynamics (screening\,/\,antisceening effects, off-Cartan fluctuations, and thermal ground-state contribution excluded), and $N_{\rm eff}$ is the effective 
number of massless neutrino flavours. As in \cite{Hahn:2018dih} we set $N_{\rm eff}$ equal to its 2015 Planck value \cite{Planck:2015fie}: $N_{\rm eff}\text{\;$=$\;}3.046$.  

The postulate SU(2)$_{\rm CMB}$ affects the comoving sound horizon $r_s(z)$, whose value at recombination 
(baryon drag) is 
the anchoring scale for the analysis of large-scale structure based on BAO, not only directly 
via the Hubble parameter $H(z)$ but also indirectly via the sound velocity $c_s$ of the 
baryon-Yang-Mills plasma conventionally modelled in terms of baryons interacting via photons. In general, $r_s(z)$ is given as 
\eqb
\label{rsz}
r_s(z)\equiv\int_z^\infty dz^\prime\,\frac{c_s(z^\prime)}{H(z^\prime)}\,,
\eqe 
where the sound velocity 
is represented by
\eqb
\label{soundcs}
c_s(z)\equiv\frac{1}{\sqrt{3(1+R(z))}}\,.
\eqe 
In what follows the subscript ${\LC}$ refers 
to the quantity computed in $\Lambda$CDM. Specifically, the ratio $R_\LC$ relates to 
entropy densities $s_\LC$ or energy densities $\rho_\LC$ of baryons (b) and photons ($\gamma$) as  
\eqb
\label{defRLam}
R_{\LC}\equiv\frac{s_{\LC,b}(z)}{s_{\LC,\gamma}(z)}\text{\;$=$\;}\frac34\frac{\rho_{\LC,b}(z)}{\rho_{\LC,\gamma}(z)}\quad (z \text{\;$\gg$\;}1)\,.
\eqe  
The generalisation of Eq.\,(\ref{defRLam}) to the baryon-Yang-Mills 
plasma replaces $s_{\LC,\gamma}(z)$ or $\rho_{\LC,\gamma}(z)$ by $s_{\SU}(z)$ or $\rho_{\SU}(z)$, 
respectively, and $s_{\LC,b}$ or $\rho_{\LC,b}(z)$ by $s_{\SU,b}$ or $\rho_{\SU,b}(z)$, respectively, to define $R_{\SU}$, see \cite{Hahn:2018dih}. 

For the epoch of recombination the postulate SU(2)$_{\rm CMB}$ predicts a significantly higher redshift 
than $\Lambda$CDM does. Namely, equating the temperature of both models at $T\text{\;$\gg$\;}T_0$, using Eq.\,(\ref{TzT>>T0}) 
for SU(2)$_{\rm CMB}$ and 
$T/T_0\text{\;$=$\;}z+1$ for $\Lambda$CDM, we arrive at 
\eqb
\label{convz}
z_{\LC}\text{\;$=$\;}\left(\frac{1}{4}\right)^{1/3}z_{\SU}\,.
\eqe 
In particular, this yields 
\eqb
\label{zeqsu2}
z_{{\SU},{\rm rec}}\text{\;$=$\;}1730\,,
\eqe
based on $z_{{\LC},{\rm rec}}\text{\;$=$\;}1090$ \cite{Planck:2015fie,Planck:2018vyg}. Repeating the argument of \cite{Hahn:2018dih}, 
we now infer from Eq. \eqref{zeqsu2} a dramatic reduction of the matter density 
parameter $\Omega_{{\SU},m,0}$ during the epoch of recombination in SU(2)$_{\rm CMB}$ compared to $\Lambda$CDM. For this purpose it is entirely sufficient to describe recombination in terms of thermodynamics (Saha approximation). The Thomson scattering rate $\Gamma$ then is a function of the recombination temperature 
$T_{\rm rec}$ only: $\Gamma\text{\;$=$\;}\Gamma(T_{\rm rec})$. Note that $T_{\rm rec}$ is independent of any cosmological model as long as thermodynamics prevails. Moreover, 
the Hubble parameter $H$ depends on $T_{\rm rec}$ via $z_{\rm rec}$: 
$H(z_{\rm rec})\text{\;$=$\;}H(z(T_{\rm rec}))$. The additional assumption, that $H$ is matter dominated during recombination turns out to be selfconsistent, see \cite{Hahn:2018dih}. Eliminating $\Gamma$ from the decoupling conditions in both models, $H_{\SU}\left(z_{\SU,{\rm rec}}\right)\text{\;$=$\;}\Gamma(T_{\rm rec})\text{\;$=$\;}H_{\LC}\left(z_{\LC,{\rm rec}}\right)$, we thus conclude that 
\eqb
\label{Omegacon}
\Omega_{\LC,m,0} \approx 4\,\Omega_{{\SU},m,0}\,.
\eqe
The most economic way for the modified cosmological model to simultaneously obey the postulate SU(2)$_{\rm CMB}$ globally and mimick $\Lambda$CDM at low redshifts\footnote{The success of $\Lambda\text{CDM}$ as a low-$z$ model is suggested by the agreement of its parameter values when extracted from purely local and different cosmology probes, see \cite{Abdalla:2022yfr}.} is the instantaneous emergence of dark matter (edm) from dark energy at some redshift $z_p<z_{\SU,{\rm rec}}$. From now on we set $z\equiv z_{\SU}$. Therefore, the following density parameter for the 
dark sector (ds) was proposed in \cite{Hahn:2018dih}:
\begin{equation}
\label{edmdef}
\Omega_{\rm ds} (z)\text{\;$=$\;}\Omega_{\Lambda} + \Omega_{\rm pdm,0} (z+1)^3 + \Omega_{\rm edm,0} \left\{  \begin{array}{lr}
\left(z_{\phantom{p}} + 1\right)^3\,, &  (z < z_p)\\
\left(z_p + 1\right)^3 \,, & (z \geq z_p)
\end{array} \right.\,.  
\end{equation}
In Eq.\,(\ref{edmdef}) today's density parameters for dark energy and dark matter are denoted by $\Omega_{\Lambda}$ and $\Omega_{\rm pdm,0}+\Omega_{\rm edm,0}\equiv \Omega_{\rm cdm,0}$, respectively, $\Omega_{\rm pdm,0}$ refers to primordial dark matter for all $z$, and $\Omega_{\rm edm,0}$ associates with dark matter emergent from dark energy at $z_p$. 
In the following a brief discussion of the physics, potentially responsible for the dark-sector model in Eq.\,(\ref{edmdef}), is given following reference \cite{MeinertHofmann2021}. 
There the dark sector starts out at the Big Bang with four species of dark energy three of which have undergone transitions into dark matter in the past; 
one species yet is to face such a transition and therefore plays the role of a cosmological constant at present. 
The theoretical underpinning of such a dark-sector model is the invocation of the axial anomaly by SU(2) Yang-Mills theories, subject to a universal Planckian Peccei-Quinn scale \cite{Giacosa:2008rw}. 
Three out of four theories presently are in confining phases with their Yang-Mills scales relating to the masses of charged leptons. Such a link to particle physics is based on the assertion that lepton doublets are 
emergent phenomena in pure SU(2) Yang-Mills theories \cite{Hofmann:2017bdz,Hofmann:2020yhn}. The associated {\sl axion} particles receive their masses $m_a$ via the axial anomaly \cite{Adler:1969er,Adler:1969gk,Bell:1969ts,Fujikawa:1979ay,Fujikawa:1980eg} invoked by topological 
charges residing in the ground states of these Yang-Mills theories and are ultralight. With a universal Planckian Peccei-Quinn scale axion masses $m_a$ thus scale like the squares of charged lepton masses $m$, e.g. ${m_{a,\mu}}/{m_{a,{\rm e}}}$\text{\;$=$\;}${m_\mu^2}/{m_{\rm e}^2}$ \cite{MeinertHofmann2021}. 

A depercolation transition from a homogeneous, superhorizon sized axion condensate (dark energy) towards a gas of nonrelativistic lumps (cold dark matter) of fuzzy dark matter (condensate core / soliton plus Navarro-Frenk-White halo) \cite{Sin:1992bg,Ji:1994xh,Matos:1998vk,Schive:2014dra,Niemeyer:2019aqm} occurs when the Hubble radius $r_H$ matches the Bohr radius $r_B$ modulo a phenomenologically determined, multiplicative constant $\alpha_e\sim55,500$, compare with \cite{MeinertHofmann2021}. For the axion particle associated with 
the electron this depercolation transition is parameterised in Eq.\,(\ref{edmdef}) to occur at $z_p\text{\;$=$\;}z_{p,e}\text{\;$=$\;}53$. 
The two other depercolation transitions, associated with the muon and the tau, are found to occur at 
$z_{p,\mu}\text{\;$=$\;}40,000$ and $z_{p,\tau}\text{\;$=$\;}685,000$ in \cite{MeinertHofmann2021}, respectively. Because the Hubble radius at $\tau$-lump depercolation is $r_H(z_{p,\tau}\text{\;$=$\;}685,000)\sim1.36\times 10^{-6}\,$Mpc this corresponds to a lower comoving cutoff scale of 0.93\,Mpc for the linear density contrast generated by adiabatic curvature perturbations. For $\mu$-lump depercolation we have $r_H(z_{p,\mu}\text{\;$=$\;}40,000)\sim3.74\times 10^{-4}\,$Mpc, corresponding to a comoving cutoff scale of 14.94\,Mpc. These two cutoff scales are well inside the nonlinear regime \cite{Miyatake:2021sdd}. For e-lump depercolation $r_H(z_{p,e}\text{\;$=$\;}53)\sim16.48$\,Mpc is obtained, associated with a comoving cutoff scale of 873\,Mpc. 
Therefore, the assumption made in \cite{MeinertHofmann2021} that density perturbations in the e-lump gas are triggered by those of the $\tau$-lump and $\mu$-lump gases is consistent for comoving scales up to 873\,Mpc. 
Beyond this scale e-lump density perturbations are seeded by adiabatic curvature perturbations upon their horizon entry.

\subsection{Cosmological parameters: \texorpdfstring{SU(2)$_{\rm CMB}$}{SU(2)CMB} vs. local and global observations in
 \texorpdfstring{$\Lambda$CDM}{ΛCDM}
\label{Cosmoparams}}

The spatially flat, global cosmological model, minimally implied by SU(2)$_{\rm CMB}$ as outlined in Sec.\,\ref{Tzoi}, and flat $\Lambda$CDM, considered as a globally valid cosmological model,  
produce the parameter values in the table below when fitted to 2015 Planck data \cite{Hahn:2018dih}, for the corresponding TT power spectrum see appendix A.
For completeness we also quote the values of flat $\Lambda$CDM fitted to 2018 Planck data \cite{Planck:2018vyg}:

\begin{table}[ht]
\caption{Best-fit cosmological parameters of flat SU(2)$_{\rm CMB}$ to the data 
in \protect\cite{Planck:2015fie} (1st column) as well as flat $\Lambda$CDM model to the data 
in \protect\cite{Planck:2015fie}, employing the TT,TE,EE+lowP+lensing likelihoods (2nd column) and to the data
in \protect\cite{Planck:2018vyg}, employing the TT,TE,EE+lowE+lensing likelihoods (3rd column). 
For SU(2)$_{\rm CMB}$ the HiLLiPOP+lowTEB+lensing likelihood is used as defined in 
\protect\cite{Planck:2015bpv} (lowP and lowTEB are pixel-based likelihoods). 
The upper section of the table quotes free parameter values, the lower 
section states the values of derived parameters. Errors correspond to 68\%-confidence levels.}

\resizebox{\columnwidth}{!}{
\begin{tabular}{c*{5}{>{}c<{}}}
 \hline
Parameter & SU(2)$_{\rm CMB}$ & $\Lambda$CDM (2015) & $\Lambda$CDM (2018) \\ [0.5ex] 
\hline
$\omega_\text{b,0}$\dotfill & $0.0173 \pm 0.0002$ & $0.02226 \pm 0.00016$ & $0.02237 \pm 0.00015$\\ 
$\omega_\text{pdm,0}$\dotfill & $0.113 \pm 0.002$ & $-$ & $-$ \\ 
$\omega_\text{edm,0}$\dotfill & $0.0771 \pm 0.0012$ & $-$ & $-$\\ 
$100\, \theta_*$\dotfill & $1.0418 \pm 0.0022$ & $1.0408 \pm 0.00032$ & $1.04092 \pm 0.00031$\\
$\tau_\text{re}$\dotfill & $0.02632 \pm 0.00218$ & $0.063 \pm 0.014$ & $0.0544 \pm 0.0073$\\
$\ln(10^{10}A_s)$\dotfill & $2.858 \pm 0.009$ & $3.059 \pm 0.025$ & $3.044 \pm 0.014$\\
$n_s$\dotfill & $0.7261 \pm 0.0058$   &$0.9653 \pm 0.0048$ & $0.9649 \pm 0.0042$ \\ 
$z_{p}$\dotfill & $52.88\pm 4.06$ &$-$ & $-$\\ [0.5ex]
 \hline
$H_0/${\tiny{km\,s$^{-1}$Mpc$^{-1}$}} \dotfill & $74.24 \pm 1.46$ &$67.51\pm 0.64$ & $67.36 \pm 0.54$\\ [0.5ex]
$z_\text{re}$\dotfill & $6.23 ^{+0.41}_{-0.42}$  & $8.5^{+1.4}_{-1.2}$ & $7.67 \pm 0.73$\\
$z_*$\dotfill & $1715.19 \pm 0.19$  & $1090.00 \pm 0.29$ & $1089.92 \pm 0.25$ \\
$z_{d}$\dotfill & $1640.87 \pm 0.27$   & $1059.62 \pm 0.31$ & $1059.94 \pm 0.30$\\
$\omega_\text{cdm,0}$\dotfill & $0.1901 \pm 0.0023$  & $0.1193\pm 0.0014 $ & $0.1200\pm 0.0012 $\\ 
$\Omega_\Lambda$\dotfill & $0.616 \pm 0.006$  & $0.6879\pm 0.0087$ & $0.6847 \pm 0.0073$\\
$\Omega_{\text{m,0}}$\dotfill & $0.384 \pm 0.006$  & $0.3121\pm 0.0087$ & $0.3153 \pm 0.0073$\\
$\sigma_8$\dotfill & $0.709 \pm 0.020$  & $0.8150\pm 0.0087$ & $0.8111 \pm 0.0060$\\
$S_8 \equiv  \sigma_8 \sqrt{\Omega_{\text{m,0}} / 0.3}$\dotfill & $0.802 \pm 0.029$  & $0.8313 \pm 0.0176$ & $0.8315 \pm 0.0137$\\
Age$/${\tiny{Gyr}}\dotfill & $11.91 \pm 0.10$ & $13.807 \pm 0.026$ & $13.797 \pm 0.023$\\ [0.5ex]
\hline\\
\end{tabular}\label{ComparisonLCDMandSU2Table}
}
\end{table}

As the table indicates, there are statistically significant deviations between 
flat SU(2)$_{\rm CMB}$ and flat $\Lambda$CDM, most noticeably in $H_0$. This $\sim4.6$ to $4.7$ $\sigma$ discrepancy is comparable 
to the one extracted from the Hubble diagram in {\sl local} flat $\Lambda$CDM, see e.g. \cite{Riess:2021jrx}, using calibrated 
standard candles, or from strong-lensing time delays (cosmography, only astrophysics model dependent extraction of $H_0$), see \cite{Wong:2019kwg}. On the other hand, fits of flat $\Lambda$CDM to BAO (standard ruler) and 2015 Planck data \cite{BOSS:2016wmc} yield a value of $H_0$ which is close to the fit of flat $\Lambda$CDM to the 2015 and 2018 Planck data alone: $(67.6\pm 0.5)$\,km\,s$^{-1}$Mpc$^{-1}$ vs. $(67.51\pm 0.64)$\,km\,s$^{-1}$Mpc$^{-1}$ (Planck 2015) and $(67.36 \pm 0.54)$\,km\,s$^{-1}$Mpc$^{-1}$ (Planck 2018), respectively. All cosmological parameters are $\sim1\,\sigma$ consistent in flat $\Lambda$CDM (2015) and flat $\Lambda$CDM (2018). 

Let us now discuss baryon density $\omega_\text{b,0}$. Global fits of flat $\Lambda$CDM to the Planck data and BBN 
yield a baryon density which is by about a factor $\sim$ 3/2 higher than the value observed by direct, local census, see e.g. \cite{Planck:2015fie,Planck:2018vyg,Kirkman:2003uv} for the former and \cite{Shull:2011aa,2019} for the latter claim. The significance of this deviation is about 2\,$\sigma$. The same tendency of such a discrepant value of $\omega_\text{b,0}$ is seen in the table when comparing flat SU(2)$_{\rm CMB}$ and flat $\Lambda$CDM, albeit at a higher significance. 

Next, in $\Lambda$CDM weak gravitational lensing effects persistently indicate a value of the clustering amplitude \cite{Miyatake:2021sdd,DES:2021wwk,Heymans:2020gsg}, characterised by $\sigma_8\text{\;$=$\;}0.718^{+0.044}_{-0.031}$ in \cite{Miyatake:2021sdd}, which is by 2-3\,$\sigma$ lower compared to its value 
extracted from CMB observation, see table and \cite{Planck:2015fie,Planck:2018vyg}. As the table indicates, 
in SU(2)$_{\rm CMB}$ the same tendency occurs, subject to a higher significance of 5.3\,$\sigma$. 
Also, there is a mild tendency for an increase of 
$\Omega_m$ in local observations (by a maximum significance of $\sim1\,\sigma$ in \cite{Miyatake:2021sdd}) compared to 
the CMB extraction in \cite{Planck:2015fie,Planck:2018vyg}. Such a tendency is also seen in the 
table, albeit now with a significance of 7.5\,$\sigma$. 

Finally, there is a $\sim2\,\sigma$ tension in the 
redshift $z_{\rm re}$ for reionisation between direct observation using the Gunn-Peterson trough 
\cite{Becker_2001} and extraction of $z_{\rm re}\text{\;$=$\;}7.67\pm 0.73$ from the 2018 Planck data  \cite{Planck:2018vyg}. 
For the 2015 Planck data, $z_{\rm re}\text{\;$=$\;}8.5^{+1.4}_{-1.2}$ this tension is 
again at $\sim2\,\sigma$. From the table we see that the tension between flat SU(2)$_{\rm CMB}$ and 
flat $\Lambda$CDM is 1.6\,$\sigma$ (2015 Planck data) and 2\,$\sigma$ (2018 Planck data).

It is conspicuous that the {\sl global} flat model SU(2)$_{\rm CMB}$ yields key cosmological parameter values which agree better with those of {\sl local} flat $\Lambda$CDM extractions rather than those of {\sl global} flat $\Lambda$CDM fitted to the same Planck data. Notice the unusually low value of the spectral index $n_s$ of adiabative curvature perturbations in SU(2)$_{\rm CMB}$, indicating a red-tilted spectrum. This may turn out to be an 
artefact of velocity divergence being suppressed on smaller scales due to late-time 
axion-condensate depercolation (e-lumps) but a modelling of the transition through an instantaneous 
transmission of this perturbation from the primordial gas ($\mu$-lumps and $\tau$-lumps), 
see \cite{Hahn:2018dih}. That is, in reality the primordial spectrum may well be scale 
invariant but is fitted to be red-tilted due to missing velocity 
divergence on smaller scales. To gain more confidence in such an interpretation a thorough 
modelling of the depercolation transition in the framework of fuzzy dark 
matter (Poisson-Schr\"odinger system) is required, see e.g. \cite{Schive:2014dra}.   

\section{CMB at large angles\label{CMBla}}

\subsection{Observational situation\label{OS}}

As exhibited in Sec.\,\ref{Cosmoparams}, the global cosmological model flat $\Lambda$CDM deviates in some key parameter values 
from both local flat $\Lambda$CDM and the global flat model SU(2)$_{\rm CMB}$ (fitted to Planck data 
and determined by angular scales associated with $l>50$ \cite{Abdalla:2022yfr,Hahn:2018dih}). In addition, there are 
inadequacies at large angular scales, see, e.g. \cite{Hinshaw:1996ut,Tegmark:2003ve,Copi:2006tu,Copi:2013cya}, for missing power in the TT correlation on angular scales larger than 60$^\circ$ and the 
breaking of statistical isotropy expressed by low-$\ell$ multipole alignment in the map of CMB temperature fluctuations  \cite{Gordon:2005ai,Copi:2005ff,Copi:2011pe,Hofmann:2013rna}. 
More specifically, based on the analysis of the two satellite missions WMAP and Planck, CMB large-angle anomalies fall into one of the following categories: lack of large-angle CMB temperature correlation (sketched above), hemispherical (dipolar) power and variance asymmetries, e.g. \cite{Eriksen:2003db,Planck:2015fie,Planck:2019evm}, octopole planarity and alignment with the quadrupole, e.g. \cite{deOliveira-Costa:2003utu,Copi:2013jna,Notari:2015kla,Schwarz:2015cma}, point-parity anomaly, e.g. \cite{Kim:2010gf,Aluri:2011wv,Gruppuso:2017nap}, variation in cosmological parameters over the sky, e.g. \cite{Fosalba:2020gls,Yeung:2022smn}, and cold spot, e.g. \cite{Vielva:2003et,Cruz:2009nd,Planck:2019evm}. 

There are many attempts at explaining the CMB large-angle anomalies in the literature, ranging from a nontrivial topology of the Universe over an unusually large matter void invoking the integrated Sachs-Wolfe effect to features in the spectra of initial perturbations, see \cite{Abdalla:2022yfr} for a recent compilation of these proposals. 
Here instead we focus on a dynamical, late-time breaking of statistical isotropy which peaks at redshift $z\sim1$ and is induced by 
screened / antiscreened photon propagation in the framework of SU(2)$_{\rm CMB}$ \cite{Hofmann:2013rna}.
This effect is expected to reduce the low-$\ell$ excess in the TT power spectrum of \cite{Hahn:2017hjt}, see appendix A.

\subsection{Modified \texorpdfstring{SU(2)$_{\rm CMB}$}{SU(2)CMB} dispersion law and CMB Boltzmann solvers}

\subsubsection{Modified photon radiance in SU(2)$_{\rm CMB}$\label{GTnu}}

\begin{figure}[H]
\centering
\includegraphics[width=\columnwidth]{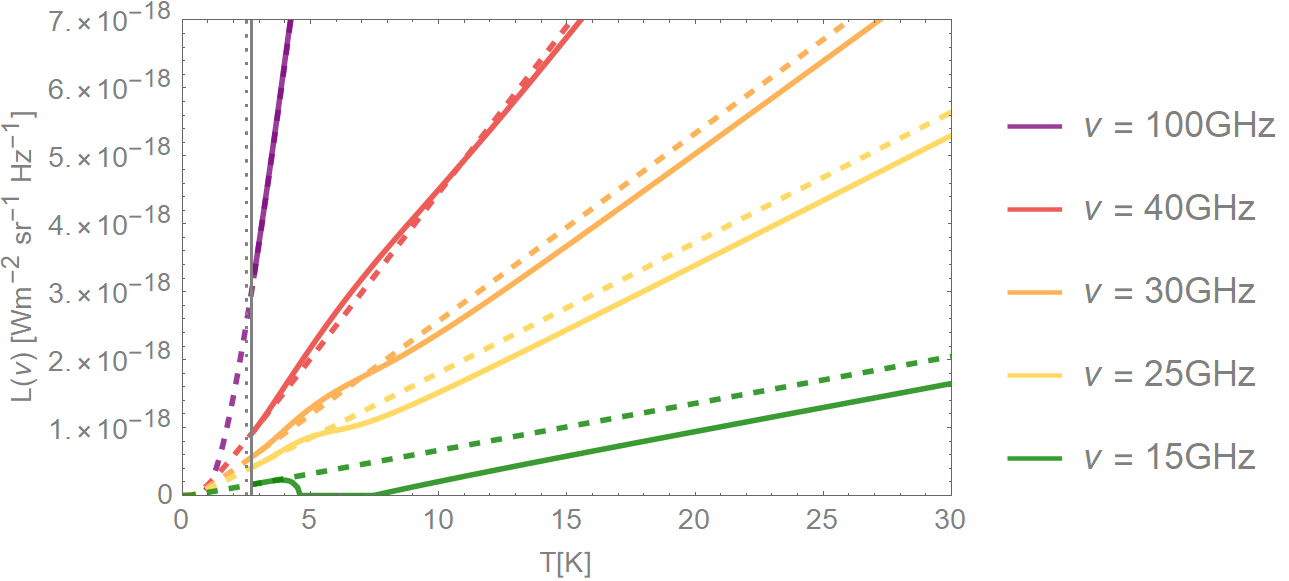}
\caption{SU(2) Yang-Mills thermodynamics exhibits three phases \protect\cite{bookHofmann}: 
the confining phase below $T_H$\,$\sim 0.9\,T_c$ (the Hagedorn temperature $T_H$ 
indicated by the vertical grey dotted line), the preconfining phase for 
$T_H$ $\le T\le T_c$ ($T_{c}$ indicated by the vertical grey line), 
and the deconfining phase for $T\text{\;$\ge$\;}T_c$. In SU(2)$_{\rm CMB}$ one has $T_c\text{\;$=$\;}T_0\text{\;$=$\;}2.725\,$K. SI units of radiance $L(\nu)$ are W\,m$^{-2}$\,sr$^{-1}$\,Hz$^{-1}$. The U(1) Rayleigh-Jeans radiances are given in dashed lines for $\nu\text{\;$=$\;}15\,$GHz (green), 25\,GHz (yellow), 30\,GHz (orange), 40\,GHz (red), and 100\,GHz (purple) while solid lines depict the associated radiances in SU(2)$_{\rm CMB}$.}
\label{thermalPhases}
\end{figure}

\begin{figure}
\centering
\includegraphics[width=\columnwidth]{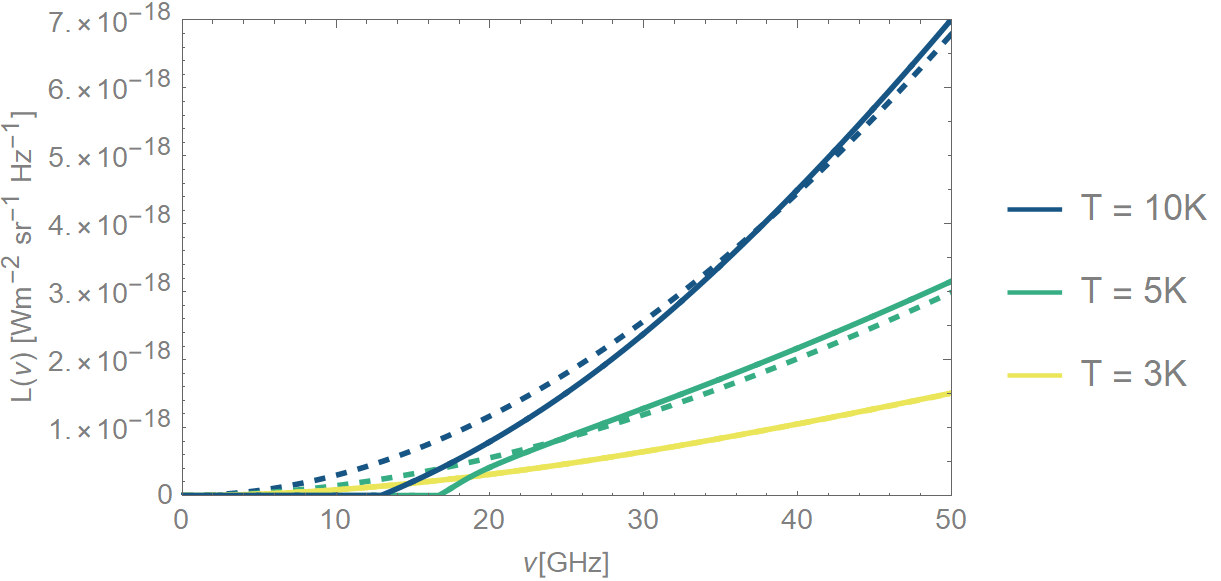}
\caption{The SU(2)$_{\rm CMB}$ and the U(1) black-body spectral radiances are shown for $T\text{\;$=$\;}3\,$K (yellow), 5\,K (green), 10\,K (blue) in solid and dashed lines, respectively. The spectral gap is widest at $\sim 2 \,\times\, T_c \approx 5\,$K. The spectral regime, where U(1) radiance is larger / smaller than SU(2)$_{\rm CMB}$ radiance, exhibits screening\,/\,antiscreening.}
\label{Gap}
\end{figure}

As explained in \cite{Schwarz:2006yz,Hofmann:2007qp}, the modified black-body spectral intensity $I_{\tiny\mbox{SU(2)}}(\nu)$ of the SU(2) theory is obtained from that of the conventional U(1) theory as follows 
\begin{equation}
\label{IntensitySU(2)}
I_{\tiny\mbox{SU(2)}}(\nu)\text{\;$=$\;}I_{\tiny\mbox{U(1)}}(\nu)\times\left(1-\frac{G(\nu)}{\nu^2}\right)\theta \left(\nu - \nu^*\right)\,,
\end{equation}
where the characteristic cutoff-frequency $\nu^*$ is defined implicitly through 
\begin{equation}
\label{nu*def}
|\vec{p}|(\nu^*)\text{\;$=$\;}\sqrt{(2\pi\nu^*)^2-G}\text{\;$=$\;}0\,,
\end{equation}
and $\theta(x)$ denotes the Heaviside function. It was shown in \cite{Falquez:2010ve} that 
\begin{equation}
\label{nu*T}
\nu^*(T)\propto T^{-1/2}\,,\ \ \ (T \text{\;$\gg$\;}T_c)\,.
\end{equation}
In SI units one has  
\begin{equation}
\label{IntensityU(1)}
I_{\tiny\mbox{U(1)}}(\nu)\equiv
\frac{2 h}{c^2}\,\nu^3\,
\text{$n_B\left(\frac{h\,\nu}{k_B T}\right)$}\,,
\end{equation}
where $k_B$ is Boltzmann's constant, $h$ is Planck's quantum of action, $c$ denotes the speed of light in vacuum, and $n_B(x)\equiv 1/(e^x-1)$. 
For the massless mode propagating into the spatial 3-direction, $\vec{p}||\mbox{e}_3$, and resorting back to natural 
units, $c\text{\;$=$\;}k_B\text{\;$=$\;}\hbar\text{\;$=$\;}1$, the screening function $G(\nu)$ 
is computed in cylindrical coordinates and reads \cite{Schwarz:2006gp} 
\begin{equation}
\label{intlimts}
\frac{G}{T^2}\text{\;$=$\;}\int d\xi\,\int d\rho\, 
e^2\lambda^{-3}\left(-4+\frac{\rho^2}{4e^2}\right)\,\rho\,
\frac{n_B\left(2\pi \lambda^{-3/2}\sqrt{\rho^2+\xi^2+4e^2}\right)}
{\sqrt{\rho^2+\xi^2+4e^2}}\,,
\end{equation}
where $\lambda\equiv 13.87\,{T}/{T_c}$ ($T_c$ the critical temperature for the deconfining-preconfining phase transition), and $e$ denotes the effective gauge coupling $e\ge\sqrt{8}\pi$. 
The support of the integration in Eq.\,(\ref{intlimts}) is determined from the demand that $\rho$ and $\xi$ satisfy one or both of the two following conditions 
\begin{equation}
\label{constcyl}
\left\lvert\frac{G}{T^2}\frac{\lambda^{3}}{(2\pi)^2}\pm\frac{\lambda^{3/2}}{\pi}\left(\sqrt{X^2+
\frac{G}{T^2}}\sqrt{\rho^2+\xi^2+4e^2}-X\xi\right)+4e^2\right\lvert\le 1\,,
\end{equation}
where $X\text{\;$=$\;}X(T,\nu)\equiv {\lvert \vec{p}\lvert}/{T}\text{\;$=$\;}{\sqrt{(2\pi\nu)^2-G}}/{T}$. For the SU(2) radiance one obtains  \cite{Falquez:2010ve}
\begin{equation}
L_{\tiny\mbox{SU(2)}}\left(T,\nu \right)\text{\;$=$\;}L_{\tiny\mbox{U(1)}} \times
                \left(1 - \frac{G}{(2\pi\nu)^2}\right)
                \theta \left( \nu - \nu^* \right)\,,
                \label{LSU2}
\end{equation}
and specifically for SU(2)$_{\rm CMB}$ one has $T_c\text{\;$=$\;}T_0\text{\;$=$\;}2.725\,$K \cite{Fixsen:2009xn,Hofmann:2009yh}. In Fig.\,\ref{thermalPhases} the temperature dependence of spectral black-body radiance in the range from $0\,-\,30$\,K is shown for five different frequencies in case of SU(2)$_{\rm CMB}$ and the conventional U(1) theory.
Notice the gap at the lowest frequency of 15\,GHz and the shifted 
linear dependence (pseudo Rayleigh-Jeans) to the right of this gap due to screening in SU(2)$_{\rm CMB}$. With increasing frequencies there is antiscreening at low temperature, which transitions into screening at higher temperatures. Both temperature regimes, screening and antiscreening, approach the U(1) radiance rapidly
as frequency increases. Fig.\,\ref{Gap} depicts the frequency dependence of spectral black-body radiance from 0 to 50\,GHz 
for three different temperatures.
It can be seen from the spectra that the 
deviation $\Delta L(\nu)\equiv |L_{\rm U(1)}(\nu)-L_{{\rm SU(2)}_{\rm CMB}}(\nu)|$ is 
maximal at $\nu\text{\;$=$\;}\nu^*$ where $\Delta L(\nu)\text{\;$=$\;}L_{\rm U(1)}(\nu)$. This is true for all temperatures. Since $L_{\rm U(1)}(\nu)\propto \nu^2\,T$ in the Rayleigh-Jeans regime and using Eq.\,(\ref{nu*T}), we conclude that 
\begin{equation} 
\label{Delta L}
\Delta L(\nu^*)\text{\;$=$\;}4.85\times 10^{-19}\,\mbox{W\,m$^{-2}$\,sr$^{-1}$\,Hz$^{-1}$}\,,\ \ \ (T \text{\;$\gg$\;}T_c)\,,   
\end{equation}
and in particular at room temperature. The feableness of such a small, maximal deviation between 
U(1) and SU(2)$_{\rm CMB}$ radiances renders the detection of the spectral 
anomaly at temperatures $T \text{\;$\gg$\;}T_c$ an experimentally challenging task.

\subsubsection{Boltzmann equation for linear perturbations of photon phase-space distribution\label{BH}}

The low-frequency, low-temperature modifications of the thermal photon dispersion law 
in SU(2)$_{\rm CMB}$ discussed in Sec.\,\ref{GTnu} imply technical difficulties in the treatment of the 
Boltzmann hierarchy for the perturbations $F_\gamma(\vec{k},\hat{n},q,\tau)$ and $G_\gamma(\vec{k},\hat{n},q,\tau)$ of the photon phase-space distribution 
in CMB codes such as CLASS \cite{Lesgourgues:2011re,Lesgourgues:2011rh}. Here $\hat{n}\equiv{\vec{q}}/{q}$.
These complications arise when evolving the latter, in comoving $\vec{k}$-space and at some comoving-momentum modulus $q$, through the low-$z$ (or large-$\tau$) regime \cite{Ma_1995}. More precisely, the low$-z$, collisionless Boltzmann equation needs to 
maintain the $q$-dependence in the perturbations $F_\gamma(\vec{k},\hat{n},q,\tau)$ 
and $G_\gamma(\vec{k},\hat{n},q,\tau)$ because the ratio 
$q/\epsilon$ ($\epsilon$ the comoving energy, see Eq.\,(\ref{ModEps}) below) depends on conformal time $\tau$. 
Here we define the perturbations (sum and difference of perturbations associated with the two independent linear polarisation states) $F_\gamma$ and $G_\gamma$ through the perturbed phase-space distribution $f$ as
\begin{equation}
f_\gamma(\vec{k},\hat{n},q,\tau)\equiv f_0(q)\big[1+F_\gamma(\vec{k},\hat{n},q,\tau)+G_\gamma(\vec{k},\hat{n},q,\tau)]\,,
\end{equation}
where 
\begin{equation}
\label{PSdistr0}
    f_0\text{\;$=$\;}f_0(\epsilon)\text{\;$=$\;}\frac{1}{4\pi^3}\frac{1}{e^{\epsilon(q,a)/T_0}-1}\,.
\end{equation}
In Eq.\,(\ref{PSdistr0}) $T_0$ is today's CMB temperature, and $a$ denotes the cosmological scale factor 
with $a(\tau_0)\text{\;$=$\;}1$ where $\tau_0$ refers to the present conformal time. In the case of thermalised photons in SU(2)$_{\rm CMB}$, a modified comoving energy-momentum dispersion law applies\footnote{Even though the dependences of $\epsilon(q,a)$ and $G(q,a)$ on scale factor $a$ and redshift $z$ are different, we abuse notation by writing $\epsilon(q,z)$ and $G(q,z)$.} \cite{Ma_1995,bookHofmann} as
\begin{equation}\label{ModEps}
    \epsilon(q,a)\text{\;$=$\;}
    \sqrt{q^2+a^2G(q,a)}\text{\;$=$\;}\sqrt{q^2+\frac{G(q,z)}{(z+1)^2}}\text{\;$=$\;}\epsilon(q,z)\,,
\end{equation}
where $G$ denotes the transverse screening function, discussed in Sec.\,\ref{GTnu} and given in Eqs.\,(\ref{intlimts}), (\ref{constcyl}), but now understood as a function of comoving momentum modulus $q$ and scale factor 
$a$ (or redshift $z\text{\;$=$\;}1/a-1$) instead of $X$ and $T$. We convert $\frac{G}{T^2}(X,T)$ of Eq.\,(\ref{intlimts}) 
to $G(q,z)$ by appealing to $X\text{\;$=$\;}{q(z+1)}/{T(z)}$ and\nn $T(z)\,/\,T_0\,=\,\mathcal{S}(z)\,(z+1)$, see Eq.\,(\ref{TzT}).\nn

\begin{figure}[H]
\centering
\includegraphics[width=\columnwidth]{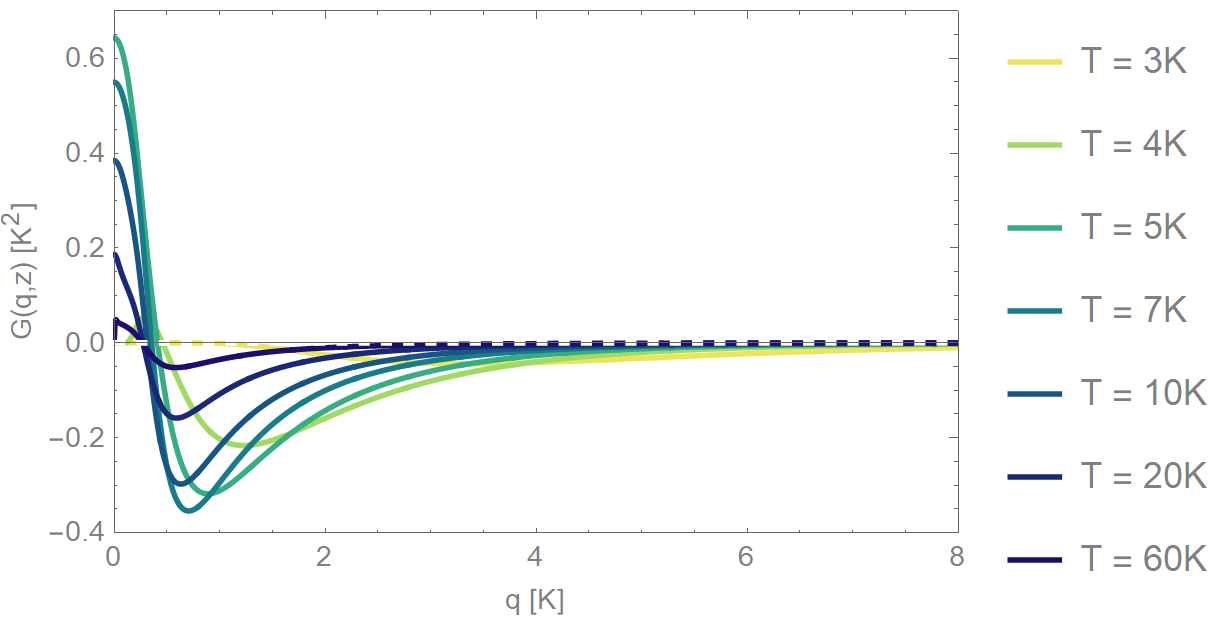}
\caption{Screening function $G(q,z)$ as a function of comoving momentum modulus $q$ and for the following temperature / redshift values:
3\,K ($z$\text{\;$=$\;}0.29, yellow), 4\,K ($z$\text{\;$=$\;}1.16, light green), 5\,K ($z$\text{\;$=$\;}1.85, green), 7\,K ($z$\text{\;$=$\;}3.07, petrol), 10\,K ($z$\text{\;$=$\;}4.83, blue), 20\,K ($z$\text{\;$=$\;}10.66, dark blue), 60\,K ($z$\text{\;$=$\;}33.96, darkest blue). The white-dashed line depicts the U(1) situation $G\equiv 0$. 
}
\label{AvgG}
\end{figure}

From Fig.\,\ref{AvgG} and Eq.\,(\ref{ModEps}) we infer that for increasing $z$ one rapidly runs into the regime of the U(1) dispersion law, $\epsilon\text{\;$=$\;}q$. In particular, the U(1) dispersion law applies prior to and through recombination.  


\begin{figure}
\centering
\includegraphics[width= \columnwidth]{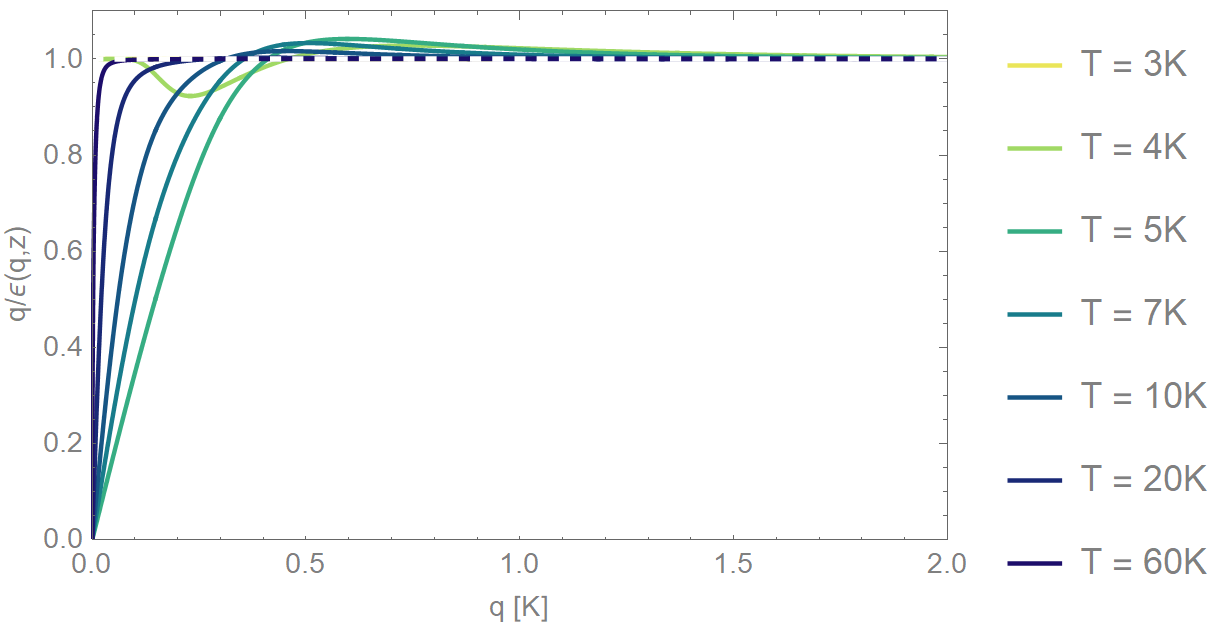}
\caption{
Shown is $q/\epsilon(q,z)$ as a function of comoving momentum $q$ and for the following temperature / redshift values: 3\,K ($z$\text{\;$=$\;}0.29, yellow), 5\,K ($z$\text{\;$=$\;}1.85, green), 7\,K ($z$\text{\;$=$\;}3.07, petrol), 10\,K ($z$\text{\;$=$\;}4.83, blue), 20\,K ($z$\text{\;$=$\;}10.66, dark blue), 60\,K ($z$\text{\;$=$\;}33.96, darkest blue). The U(1) behaviour $q/\epsilon\equiv 1$ largely coincides with the behaviours at 3\,K and 60\,K and is shown in terms of a white-dashed line.
}
\label{q/ep for avg G}
\end{figure}

In conformal Newtonian gauge, the linear perturbation $\Psi(\vec{k},\hat{n},q,\tau)\text{\;$=$\;}F_\gamma(\vec{k},\hat{n},q,\tau)+G_\gamma(\vec{k},\hat{n},q,\tau)$ evolves according to the Boltzmann equation \cite{Ma_1995}
\begin{equation}\label{Boltzmann}
    \frac{\partial \Psi}{\partial \tau} + i\frac{q}{\epsilon}(\vec{k}\cdot\hat{n})\Psi + \frac{d\ln f_0}{d\ln q}\Big[\dot{\phi} - i\frac{\epsilon}{q}(\vec{k}\cdot\hat{n})\psi\Big]\text{\;$=$\;}\frac{1}{f_0}\Big(\frac{\partial f}{\partial \tau}\Big)_C\,,
\end{equation}
where $\phi$ and $\psi$ are the gravitational potentials, and the right-hand side is the collision term. This term is only relevant prior to and through recombination and depends on $F_\gamma(\vec{k},\hat{n},q,\tau)$ and $G_\gamma(\vec{k},\hat{n},q,\tau)$ separately \cite{Ma_1995}. The expansion of $F_\gamma(\vec{k},\hat{n},q,\tau)$ and $G_\gamma(\vec{k},\hat{n},q,\tau)$ into Legendre polynomials $P_\ell(\hat{k}\cdot\hat{n})$ ($\ell\text{\;$=$\;}0,1,2,\cdots$) yields coefficients $F_{\gamma,\ell}(\vec{k},q,\tau)$ and $G_{\gamma,\ell}(\vec{k},q,\tau)$ 
which evolve in $\tau$ (or $z$) according to a Boltzmann hierarchy \cite{Ma_1995}. 
To perform a match of high-$z$ and low-$z$ (and therefore large-angle) downward evolutions at some appropriate, intermediate value $z_\text{match}$ we notice that the need to retain the $q$-dependence in $F_{\gamma,\ell}(\vec{k},q,\tau)$ and $G_{\gamma,\ell}(\vec{k},q,\tau)$ at low $z$ (not integrating it out) also requires to keep it at high $z$ where antiscreening\,/\,screening effects of SU(2)$_{\rm CMB}$ 
can safely be neglected. Therefore, the Boltzmann hierarchy needs to be solved 
on a $q$-grid for all $z$. In Fig.\,\ref{q/ep for avg G} the $z$-evolution of the factor $q/\epsilon$,
which induces this complication, is shown for low values of $z$.  

\subsubsection{Structure of CLASS}
\noindent Having (i) reviewed the main features of SU(2)$_{\rm CMB}$ as a cosmological model, see \cite{Hahn:2018dih}, 
and (ii) considered the CMB at large angular scales within deconfining SU(2) Yang-Mills thermodynamics, see \cite{Ludescher:2009my,Hofmann:2013rna}, we may now discuss what it takes to quantitatively 
confront SU(2)$_{\rm CMB}$ with the observed large-angle anomalies discussed in Sec.\,\ref{OS}. 

Several CMB Boltzmann codes are available such as CMBFAST \cite{Seljak_1996}, CMBEASY \cite{Doran_2005}, CAMB \cite{Lewis_2000}, and CLASS \cite{Lesgourgues:2011re}. Here we 
choose to discuss CLASS due to its flexibility, speed, and good documentation which also has motivated its use in \cite{Hahn:2018dih}. 

\begin{figure}\centering
\includegraphics[width=\columnwidth]{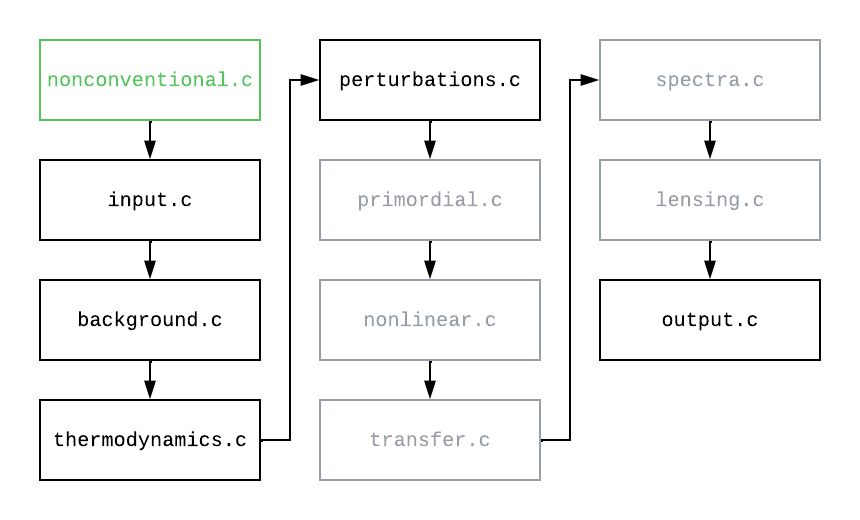}
\caption{Modules of CLASS. The module {\sl nonconventional.c} was introduced in \protect\cite{Hahn:2018dih} into the backbone of CLASS. The modules in black contain SU(2)$_{\rm CMB}$ modifications whereas the modules in gray are untouched.}
\label{flowchartmodules}
\end{figure}

CLASS is written in pure C and includes the following modules: {\sl input.c}, {\sl background.c}, {\sl thermodynamics.c}, {\sl perturbations.c}, {\sl primordial.c}, {\sl nonlinear.c}, {\sl transfer.c}, {\sl spectra.c}, {\sl lensing.c}, and {\sl output.c}. Each of these modules perform specific tasks and feed their outputs into the subsequent module along the aforementioned order.  
The following modifications were implemented in \cite{Hahn:2018dih}:\\ 
\noindent (i) A module called {\sl nonconventional.c} was added which computes the thermodynamical quantities $\rho_{\SU}$ (energy density), $P_{\SU}$ (pressure), and the scaling function $\mathcal{S}(z)$ of Eqs.\,(\ref{devlinS}) and \eqref{TzT} in SU(2)$_{\rm CMB}$.\\ 
\noindent (ii) The module {\sl input.c} contains all input and precision parameters. Additional cosmological parameters in SU(2)$_{\rm CMB}$ such as $z_p$, $\Omega_\text{edm,0}$ and the new conversion between the neutrino temperature $T_\nu$ and the CMB temperature $T$, see Eq.\,(\ref{eq:def:omegaNu}), are introduced here.\\  
\noindent (iii)  The module {\sl background.c} solves the Friedmann equation and stores other quantities such as the energy densities of individual species ($\rho_i$), the critical density ($\rho_{c}$), the Hubble parameter $H$, and conformal time $\tau$. Within this module, the new cosmological model is implemented according to Sec.\,\ref{Tzoi}. Also, the ratio $R_{\SU}\equiv{\frac{s_b(z)}{s_{\SU}(z)}\text{\;$=$\;}\frac{3}{4}\frac{\rho_b(z)}{\rho_{\SU}(z)}}$, see Eq.\,(\ref{defRLam}) and text following it, is defined here.\\  
\noindent (iv) The module {\sl thermodynamics.c} evolves the baryon-photon plasma, relying on the modified sound speed $c_s(z)\equiv 1/\left({3(1+R_{\SU}(z))}\right)$, and stores quantities such as the ionisation fraction $\chi_e$ as well as recombination and reionisation redshifts. The modified $T-z$ relationship for SU(2)$_{\rm CMB}$ in Eq.\,\eqref{TzT} is implemented within this module.\\ 
\noindent (v) The module {\sl perturbations.c} solves the perturbations evolution for each specified particle species and gravity. This module also includes the Euler equation for the emergent dark matter component \cite{Hahn:2018dih}.\\  
\noindent (vi) The {\sl output.c} module is extended to include the new SU(2)$_{\rm CMB}$ parameters.\\ 
\noindent All other modules, namely {\sl primordial.c}, {\sl nonlinear.c}, {\sl transfer.c}, {\sl spectra.c}, and {\sl lensing.c} are not directly affected by SU(2)$_{\rm CMB}$. Fig.\,\ref{flowchartmodules} provides an overview on CLASS modules, how they depend on one another, and which 
modules are modified / added in \cite{Hahn:2018dih} due to SU(2)$_{\rm CMB}$.
\noindent 
To address large-angle anomalies via a modification in the hierarchy for perturbations of the 
photon phase-space distribution, as discussed in Sec.\,\ref{BH}, we foresee the following changes in the SU(2)$_{\rm CMB}$ modified version of CLASS:\\
\noindent (A) We are required to set up a function in {\sl class.c} which calls the screening function $G(q,z(\tau))$ for a given comoving momentum $q$ and redshift $z$ (and thus conformal time $\tau$) from precomputed tables. \\ 
\noindent  (B) In the {\sl perturbations.c} module, the evolution of the perturbations of the photon phase-space distribution needs to be performed on a 
$q$-grid along an $\ell$-hierarchy (similar to the non-cold dark matter species description in \cite{Lesgourgues:2011rh}). This also requires the introduction of functions $\epsilon$ (Eq.\,\eqref{ModEps}) and $q/\epsilon$ as well as the corresponding modification of 
$\frac{d\ln f_0}{d\ln q}$ in Eq.\,(\ref{Boltzmann}). \\ 

\noindent The bottleneck is the implementation of the $q$-grid in the Boltzmann hierarchy for 
$F_{\gamma,\ell}$ and $G_{\gamma,\ell}$ which also involves the collision terms that are active prior to 
and throughout recombination (high-$z$ case). Moreover, a matching\footnote{Matching at $z_\text{match}$ means that for $z\text{\;$\ge$\;}z_\text{match}$ 
we set $G\equiv 0$ in the Boltzmann hierarchy while $G$ is taken from a precomputed table for $z<z_\text{match}$.} at some intermediate $z_\text{match}$ with $1\ll z_\text{match}\ll z_*$ of high-$z$ and low-$z$ evolutions needs to implemented in {\sl perturbations.c} on the $q$-grid such that power spectra do not 
depend on the choice of $z_\text{match}$.

It is hoped that a group with good experience in the implementation of the hierarchy 
for massive neutrinos in CLASS may be interested in pursuing the above mentioned code modifications, 
desirably in collaboration with the present authors. 

\section{Summary and Outlook}

In the present paper we have reviewed the cosmological model SU(2)$_{\rm CMB}$ which assumes 
that the CMB is subject to deconfining SU(2) Yang-Mills thermodynamics and a spatially 
flat Universe. This model coincides with flat $\Lambda$CDM locally. However, due to a modified 
temperature ($T$) - redshift ($z$) relation SU(2)$_{\rm CMB}$ deviates strongly from flat $\Lambda$CDM at high $z$  
with profound implications for the dark-sector physics. Cosmological parameter values, which are not affected by low-$z$ physics or the low-$\ell$ multipoles,  were extracted in \cite{Hahn:2018dih} by fits to 2015 Planck power spectra. The corresponding model then yields an excess of low-$\ell$ 
power in TT which we address in the second part of the present paper in terms of photon screening\,/\,antiscreening effects at low $z$. Here, we confirm that 
there is indeed no influence of these effects on cosmological parameter fitting. We have compared the according parameter values 
of SU(2)$_{\rm CMB}$ with extractions from cosmologically local and global data within flat $\Lambda$CDM or within a 
cosmographic context. As a result, we see a tendency of SU(2)$_{\rm CMB}$ as a global model to lean 
towards locally extracted cosmological parameter values of $H_0$, $z_{\rm re}$, $\omega_b$, $\sigma_8$, 
and $\Omega_m$. 

The low-$z$ spectral radiance antiscreening\,/\,screening anomalies in the Rayleigh-Jeans regime of deconfining SU(2)$_{\rm CMB}$ thermodynamics were not considered in \cite{Hahn:2018dih} but are expected to relate to the large-angle anomalies of the CMB \cite{Ludescher:2009my,Hofmann:2013rna} and to induce a lowering of low-$\ell$ power in TT, see Fig.\,\ref{2015fit} in appendix A.
Their implementation in CMB Boltzmann codes is arduous because of the need 
to introduce a $q$-grid for the Boltzmann hierarchy of perturbations to the photon phase-space 
distribution and a match of low-$z$ with high-$z$ evolutions. Hoping that groups more experienced 
with the implementation of Boltzmann codes for massive, relativistic species may be interested 
in pursuing an SU(2)$_{\rm CMB}$ code modification, desirably together with the present authors, we have provided information on 
the low-$z$ dependence of the screening function $G$ and the associated 
modified comoving energy-momentum relation for the photon. We have also discussed which CLASS 
modules need to be targeted in implementing SU(2)$_{\rm CMB}$ modifications, both for the cosmological model \cite{Hahn:2018dih} and 
the linear perturbations thereof. 

If the SU(2)$_{\rm CMB}$ modifications of CMB codes proposed in the present paper turn out to yield the 
lowering of low-$\ell$ power in TT under the assumption of statistical isotropy in projecting 
onto the C$_\ell$'s, this would motivate a dedicated analysis of statistical isotropy breaking in terms of less inclusive statistics as a next step. Also, as discussed in Sec.\,\ref{Cosmoparams}, a modelling of the depercolation transition from dark energy to dark matter, using the framework of fuzzy dark matter from ultralight axions, would refine Eq.\,(\ref{edmdef}) 
and yield insights in nonlinear structure formation on small scales \cite{Miyatake:2021sdd}.

\section{Data availability}
\noindent The {\sl Mathematica} notebooks for the computation of the screening function $G$ and the 
coefficient $q/\epsilon$ as well as the modified CLASS code of \cite{Hahn:2018dih} are 
available from the authors upon request. 

\section{Acknowledgements}
\noindent JM's work is supported by the Vector Foundation under grant number P2021-0102. The authors would like to acknowledge useful discussions with Daniel Kramer and Philip Matthias.


\bibliographystyle{mnras}
\bibliography{BibSU2CMB}

\appendix

\newpage
\section*{Appendix A: \texorpdfstring{SU(2)$_{\rm CMB}$}{SU(2)CMB} CMB fit}\label{Appendix A}

\noindent A lowering of the $TT$ power spectrum for small $\ell$ is expected \cite{Hofmann:2013rna} 
when taking screening\,/\,antiscreening effects into account in the comoving dispersion law of the low-$z$ photon, see Eq.\,(\ref{ModEps}) and Fig.\,\ref{q/ep for avg G}. This may close the red shaded area in Fig.\,\ref{2015fit} below. 
Beyond such a lowering of small-$\ell$ $TT$ power, the investigation of statistical-isotropy breaking, induced by a cosmologically local temperature depression and characterised by a typical gradient \cite{Gordon:2005ai,Hofmann:2013rna}, requires less inclusive statistics, see e.g. \cite{Schwarz:2015cma}.

\begin{figure}[H]
\centering
\caption{\protect{\label{2015fit}}
Normalised power spectra of TT correlator for best-fit parameter values quoted in Table\,\ref{ComparisonLCDMandSU2Table}.
Figure adapted from \protect\cite{Hahn:2018dih}.}
\includegraphics[width=\columnwidth]{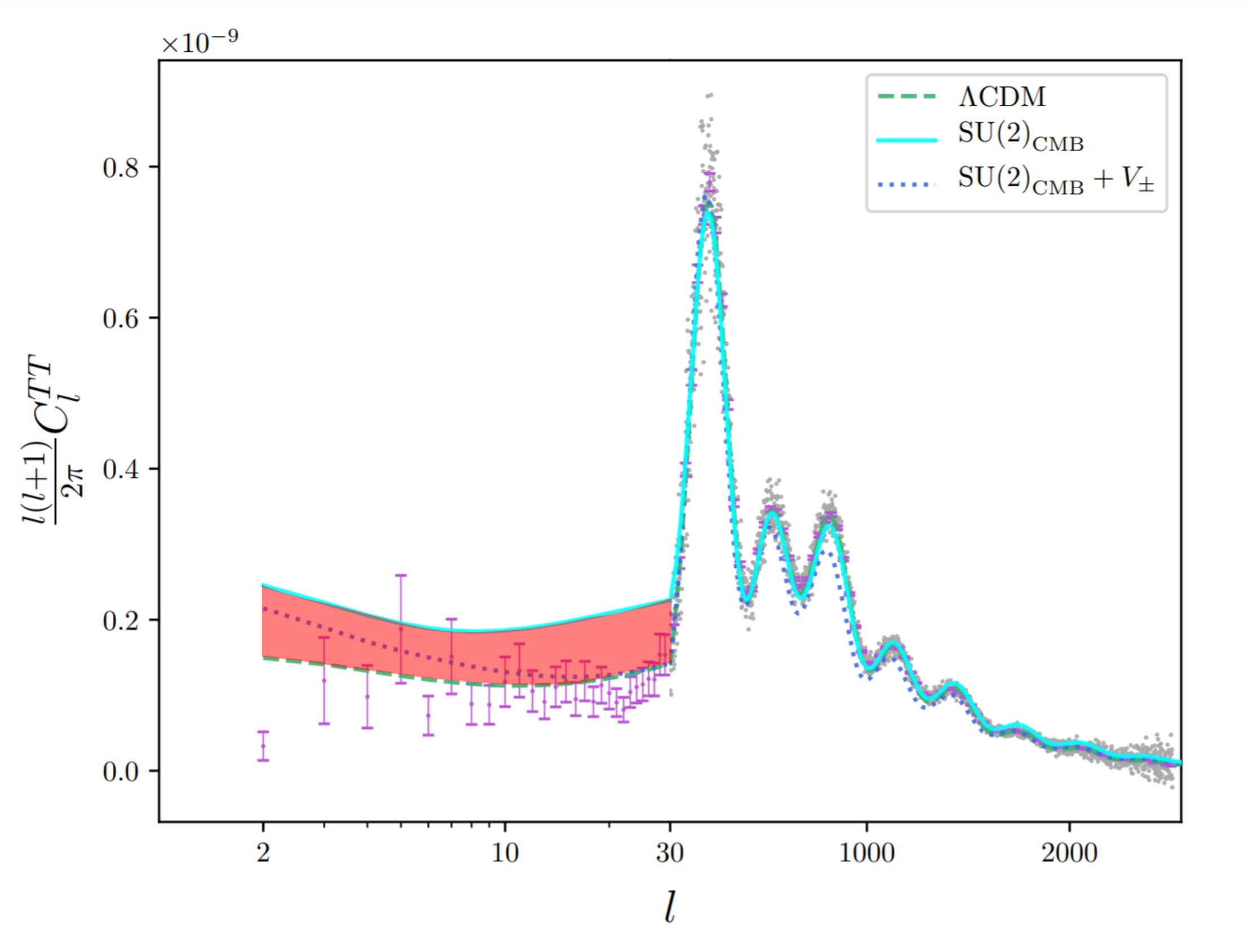} 
\end{figure}

\label{lastpage}
\end{document}